\documentstyle [12pt]{article}
{}

\def\plb#1 #2 {Phys. Lett. {\bf#1B} #2 }
\def\npb#1 #2 {Nucl. Phys. {\bf B#1} #2 }
\def\prl#1 #2 {Phys. Rev. Lett. {\bf #1} #2 }
\def\cmp#1 #2 {Comm. Math. Phys. {\bf #1} #2 }
\def\mpl#1 #2 {Mod. Phys. Lett. {\bf A#1} #2 }

\def\inbar{\vrule height1.5ex width.4pt depth0pt}
\def\IB{\relax{\rm I\kern-.18em B}}
\def\IC{\relax\,\hbox{$\inbar\kern-.3em{\rm C}$}}
\def\ID{\relax{\rm I\kern-.18em D}}
\def\IE{\relax{\rm I\kern-.18em E}}
\def\IF{\relax{\rm I\kern-.18em F}}
\def\IG{\relax\,\hbox{$\inbar\kern-.3em{\rm G}$}}
\def\IH{\relax{\rm I\kern-.18em H}}
\def\II{\relax{\rm I\kern-.18em I}}
\def\IK{\relax{\rm I\kern-.18em K}}
\def\IL{\relax{\rm I\kern-.18em L}}
\def\IM{\relax{\rm I\kern-.18em M}}
\def\IN{\relax{\rm I\kern-.18em N}}
\def\IO{\relax\,\hbox{$\inbar\kern-.3em{\rm O}$}}
\def\IP{\relax{\rm I\kern-.18em P}}
\def\IQ{\relax\,\hbox{$\inbar\kern-.3em{\rm Q}$}}
\def\IR{\relax{\rm I\kern-.18em R}}
\def\ZZ{\relax{\sf Z\kern-.4em Z}}

\def\beq{\begin{equation}}    \def\eeq{\end{equation}}

\def\beann{\begin{eqnarray*}} \def\eeann{\end{eqnarray*}}
\def\bea{\begin{eqnarray}}    \def\eea{\end{eqnarray}}
\def\lleq#1{\label{#1}\eeq}   

\def\fnote#1#2{\begingroup\def\thefootnote{#1}
\footnote{#2}\addtocounter
{footnote}{-1}\endgroup}

\def\notin{\ \hbox{{$\in$}\kern-.51em\hbox{/}}}

  \def\th{\theta}  
\def\bz{\bar z} \def\bth{\bar {\theta}}

 \def\cO{{\cal O}}

\def\tg{\tilde g} \def\tn{{\tilde n}} \def\tW{\widetilde W}
\def\tPhi{\widetilde {\Phi}}  \def\tPsi{\widetilde {\Psi}}

\begin{document}
\hfill
{HD--THEP--92--37}
\vskip 1.2truein
\centerline{\bf NOVEL FLOWS BETWEEN N$=$2 LANDAU--GINZBURG THEORIES:}
\vskip .1truein
\centerline{\sc New Directions in Modulii Space via $c=0$
                       Theories}

\vskip .7truein


\centerline {\sc R. Schimmrigk \fnote{\diamond}
                {\normalsize {Bitnet address: q25@dhdurz1.bitnet}}}

\vskip .3truein


\centerline {\it Institut f\"ur Theoretische Physik,
                   Universit\"at Heidelberg}
\centerline {\it Philosophenweg 16, 6900 Heidelberg, FRG}

\vskip 2truein
\centerline{\bf ABSTRACT}

\vglue 0.3cm
{\rightskip=1pc
 \leftskip=1pc
\noindent
A new method for constructing flows between distinct Landau-Ginzburg
theories at fixed central charge is presented. The essential
ingredient of the construction is an enlarged moduli space obtained
by adding theories with zero central charge. The flows involve only
marginal directions hence they can be applied to transitions between
string vacua, in particular to the construction of mirror pairs of
string ground states described by RG fixed points of N=2
supersymmetric Landau--Ginzburg theories. In contrast to previous
methods this new construction of mirror theories does not depend on
particular symmetries of the original theory.
}
\vfill
\renewcommand\thepage{}
\eject

\baselineskip=16.5pt
\parindent=21pt
\pagenumbering{arabic}
\parskip=10pt

\vskip .3truein
\noindent
{\sc 1. Introduction}

\noindent
An important question in quantum field theory is the problem of
constructing flows between renormalization group fixed points. A
landmark result in this direction is Zamolodchikov's `$c$--theorem'
[1] which shows that along the RG trajectories the $c$--function,
describing the central charge at the fixed points, decreases
monotonically.
This result unfortunately seems to imply that RG flows are
uninteresting
for string theory where the ground states are described by conformal
field theories at some fixed central charge. The only possible flows
therefore are those generated by marginal operators. But flows along
marginal directions, as long as they are regular, only change the
Yukawa couplings and do not change the spectrum of the theory. When
the flows become (in some sense) singular [2] control of the spectrum
of the theory is lost.

It is the purpose of this article to introduce a new type of flow
which relates distinct N$=$2 supersymmetric Landau--Ginzburg theories
at some fixed central charge. These flows proceed via marginal
operators,
hence are of interest for the understanding of string vacua described
by RG fixed points of Landau--Ginzburg theories. They lead, in
particular, to a new type of mirror map between pairs of string vacua.
Most known constructions of mirror theories depend, in general, on
orbifolding of the original theory [3--6]
\fnote{1}{An exception is Batyrev's construction via dual
          polyhedra [7].}.
This is unsatisfactory because no general selection rule is known
which
determines the specific form of the action or even its order. Even
though for Fermat type superpotentials both ingredients, the order
of
the group and its specific form is known [8], for non--Fermat
polynomials finding the action is more of an art. Thus it is useful
to search for constructions which do not depend on the modding of
symmetries.

In Section 2 I will describe the basic idea, keeping the discussion
general. Even though the application I will focus on in this paper
 is mirror
symmetry, the basic framework is quite independent of that particular
question and may well have other applications. Concrete applications
to various types of potentials will be presented in Section 3 and in
Section 4 I will discuss mirror pairs of Landau--Ginzburg vacua.
Section 5 contains results regarding the explicit construction of
mirror vacua via marginal flows.

\vskip .3truein

\noindent
{\sc 2. Enlarged Moduli Space via $c=0$ Theories: The Basic Idea}

\noindent
To motivate the construction consider a heterotic vacuum described by
a GSO--projected Landau--Ginzburg
theory defined by a superpotential $W(\Phi_i)$ that depends on chiral
N$=$2 superfields $\Phi_i(z,\bz,\th^{\pm},\bth^{\pm})$, $~i=1,\dots,n$.
The mirror question is how to construct the mirror theory, which I
assume
to be described by the mirror potential $\tW(\tPhi_i)$ which in turn
depends on chiral N$=$2 superfields
$\tPhi_i(z,\bz,\th^{\pm},\bth^{\pm})$, $~i=1,\dots,\tn$.
As a first step toward an explicit construction of the mirror vacuum
$\tW$ from $W$ one
may assume a candidate pair of potentials $(W,\tW)$ as given and to
attempt to somehow `move' from one theory to the other.

In order to achieve this one might imagine adding the two potentials
and performing some sort of projection in the path integral which
reduces
this `superposition' of theories to the individual potentials.
This does not work because the central charge is doubled in the
process and therefore the theory $W+\tW$ does not describe a
heterotic
vacuum of the type necessary. This problem can be circumvented
however.

Given an arbitrary N$=$2 supersymmetric Landau--Ginzburg theory
$W$ with
some central charge it is possible to, in a sense, add some new
nontrivial
theory $\tW$ without changing the central charge by doing something
that I will call `trivialization'.
The idea is to compensate the central charge of the new theory by
introducing additional fields with negative central charge
\fnote{1}{It should be emphasized that this does not mean that
          the {\it dimensions} of these additional fields are
	  negative,
          which would be a disaster.}
 such that the added theory has in fact vanishing central charge.
To find such fields and the appropriate action is easy and always
possible:
for any superpotential $\tW(\tPhi_i)$ the potential
\beq
\tW^{'}(\tPhi_i,\Psi_i) := \tW(\tPhi_i) + \sum_i g_i \tPhi_i\Psi_i
\eeq
leads to a $c=0$ theory. I will call this theory the `trivialized'
theory.
Adding this trivialized theory to the original potential $W(\Phi_i)$
therefore does not change the central charge. Furthermore the
spectrum remains the same because the chiral ring of the theory
$\tW'$ is
trivial owing to the fact that $q_{\Psi_i} = 1 - q_{\tPhi_i}$, hence
$\mu(\tW')=1$
\fnote{2}{More explicitly this can be seen by considering the ideal
          generated by the potential $\tW'$.}.

Adding the
potential $\tW^{'}(\tPhi_i,\Psi_i)$ is, however, not merely a trivial
rewriting of the original
theory because the parameter space of the new theory can contain
deformations
along `mixed' directions, involving fields of the type
$\cO_{IJ}= \Phi_I \Psi_J$, where $I,J$ are multiindices, i.e. the
operators
$\cO_{IJ}$ describe monomials in the
variables $\Phi_i$ and $\Psi_j$. The existence of these mixed
directions
is nontrivial. Assuming they exist the question arises what their
general
structure is. Since the variables with negative central charge were
introduced
via terms of the form $\tPhi_i\Psi_i$ these new operators are not
bilinear
but instead will be, in general, of the form $\Phi_i^{a_j} \Psi_j$.
Denote
the potential containing the couplings between the
 $\Phi_i$'s and the $\Psi_j$'s by $ W(\Phi_i,\Psi_j)$. It is then
 possible to move to a point in the enlarged parameter space where
 the total potential is of the form
\beq
W(\Phi_i) + W(\Phi_i,\Psi_j) + \tW(\tPhi_i),
\eeq
i.e. the $\Psi_j$ are completely decoupled from the $\tPhi_i$. At
this point
the theory becomes more singular than the original theory: the
singular
set is not just an isolated point at the origin but some higher
dimensional submanifold. This is
precisely what is needed because it allows the spectrum of the theory
to change.
If it so happens that the first two potentials define a $c=0$ theory
it is
possible to split off this new trivialized theory and be left with a
potential
that describes a different consistent Landau--Ginzburg theory
with an isolated singularity, i.e. with a finite chiral ring.

In a nutshell, then, the idea is to add to the moduli space of
Landau--Ginzburg
theories new directions by adding $c=0$ theories. Moving in this
enlarged
configuration space allows to connect theories with different spectra
by
passing through degenerate theories which can be made into regular
ones by splitting off different $c=0$ theories.

\vskip .3truein
\noindent
{\sc 3. Examples: Coset Landau--Ginzburg Theories and Others}

\noindent
To be concrete consider first a pair of potentials which derives [9,10]
 from a minimal N$=$2 superconformal theory at level $k$ with central
charge $c=3k/(k+2)$; the first one
\beq
W_k(\Phi) = \Phi^{k+2}
\lleq{minidiag}
describes the mean field theory of the exactly solvable model endowed
with the diagonal affine invariant whereas the second one
\beq
\tW(\tPhi_1,\tPhi_2) = \tPhi_1^k+\tPhi_1\tPhi_2^2
\lleq{minid}
describes the theory with the nondiagonal affine D--invariant.

The extended theory
\beq
W(\Phi) + \tW'(\tPhi_i,\Psi_i) =
\Phi^{k+2}+\tPhi_1^k+\tPhi_1\tPhi_2^2 +
\tg_1 \tPhi_1\Psi_1+\tg_2 \tPhi_2\Psi_2.
\eeq
leads to an enlarged moduli space which contains mixed monomials of
charge one, e.g.
\beq
\Phi^2\Psi_1,~~~~ \Phi^{k-1}\Psi_2
\eeq
describing new marginal directions along which the original theory
can be deformed.  Thus it is possible to move to the point in
parameter space  defined by
\beq
W(\Phi) + g_1 \Phi^2 \Psi_1 + g_2 \Phi^{k-1} \Psi_2 + \tW(\tPhi_i)
\eeq
where the fields with negative central charge are completely
decoupled from $\tW(\tPhi_i)$. This theory does not have a
well--defined chiral ring.
However, the first three terms define a $c=0$ theory
which can be split off, leaving a nondiagonal theory with a
well--defined spectrum which is different from the spectrum of the
diagonal model.

It is worth remarking that there exists a framework in which the
potentials
just described appear in a natural way, namely the level--rank
duality [11] which relates at level $k$ the two theories
\beq
\frac{SU(2)}{ U(1)}~{\Bigg |}_k ~~\longleftrightarrow ~~
\frac{SU(k+1)}{SU(k)\times U(1)}~{\Bigg |}_1.
\eeq
Consider the potentials for these two theories. For any theory of
the type
\beq
\frac{SU(m+n)}{SU(m)\times SU(n)\times U(1)}~~{\Bigg |}_1
\eeq
the potential is known [12,13] to be of the form
\beq
W= \sum_{n_1+2n_2+\cdots mn_m=m+n+1} A_{n_1\dots n_m}
                        \Phi_1^{n_1} \cdots \Phi_m^{n_m} .
\eeq
Thus the potential of the minimal model is
\beq
W = \Phi_1^{k+2} + \Phi_2^2
\eeq
whereas the potential of the dual model is in the deformation class
\fnote{3}{Note that this potential, as written, appears to describe
the
          sum of two trivial theories with $c=0$. This is not the
	  case
	  however, since the potential in fact does not lead to a
	  finite
	  chiral ring at all: the theory is singular.}
\beq
\tW= \Phi_1^{k+2} + \Phi_2^{(k+2)/2} + \Phi_2\Phi_{k/2}^2 +
\Phi_{(k+2)/2}^2
  + \Phi_2 \Phi_k + \Phi_{k/2} \Phi_{k/2 +2}
  + \Phi_1^2 \Phi_k + \Phi_1^{k/2} \Phi_{k/2+2} + \cdots
\eeq
It is therefore always possible to split off a trivial $c=0$ theory
in the
level--rank--dual theory to be left with a minimal theory in which
the diagonal affine invariant has been replaced by the D--invariant.

With the Landau--Ginzburg theories just described it is possible to
analyze
a large variety of mirror pairs by tensoring other exactly solvable
N$=$2 models or, more generally, by adding some arbitrary N$=$2
supersymmetric Landau--Ginzburg potential. For the sake of generality
it should be
emphasized however that the idea introduced in Section 2 can be
applied to
Landau--Ginzburg potentials that do not derive from the coset
construction.
A simple example may illustrate how an iterative application of the
technique allows to discuss other types of theories as well.
Consider the pair of potentials described by
\beq
W(\Phi_i) = \Phi_1^{25} + \Phi_1\Phi_2^{16}
\lleq{minidiagit}
and
\beq
\tW(\tPhi_i) = \tPhi_1^{25} + \tPhi_1\tPhi_2^8 + \tPhi_2\tPhi_3^2,
\lleq{minidit}
the latter leading to the $c=0$ theory
\beq
\tW'(\tPhi_i, \Psi_i) =\tW(\tPhi_i) + \sum \tPhi_i\Psi_i.
\eeq
Moving along mixed direction it is possible to arrive at the potential
\beq
W(\Phi_i)+\Phi_1\Psi_1 +\Phi_2^2\Psi_2 + \Phi_1^{11}\Psi_3 +\tW(\tPhi_i).
\eeq
After splitting off the $c=0$ theory defined by the first four terms
one is left with the new theory $\tW(\tPhi_i)$.

It is easy to see that the example of the last paragraph is just an
iteration
of the basic construction involving the transition from the diagonal
affine invariant of a minimal model at level $(2b-2)$,
$W(\Phi_2) = \Phi_2^{2b}$, to the nondiagonal D--invariant
$W(\Phi_i) = \tPhi_2^{b} +\tPhi_2\tPhi_3^2$. The only difference is
that the fields which carry the flow are factored into some additional
fields.
 Similarly one can consider further iterations to construct flows
 between ever more complicated potentials.

There are, however, other potentials that occur in  the class of all
Landau--Ginzburg vacua [14,15] which are different from the types
described so far. Even though it is not, at present, known what
precisely
the range of applicability of this new construction is, it is clear
that
it is
more general than the infinite series discussed above. An interesting
example is the following pair which involves the transposition of a
potential, a concept introduced in ref. [6]. The example is also
enlightening because it illustrates how the precise vacuum structure
comes to the rescue in cases where a naive application of the
construction does not work.

Consider the pair of potentials described by
\beq
W(\Phi_i) = \Phi_1^5\Phi_2 + \Phi_2^3
\eeq
and
\beq
\tW(\tPhi_i) = \tPhi_1^5  + \Phi_1 \tPhi_2^3
\eeq
neither of which is related to a minimal model. The trivialization of
the potential $\tW$ introduces two fields $\Psi_1,\Psi_2$ with charges
$4/5,11/15$. Whereas it is possible to couple $\Psi_2$ to the
field $\Phi_1$, it is not possible to couple $\Psi_1$ to either
$\Phi_1$ or to $\Phi_2$.
Thus it is impossible to decouple the fields $\Psi_i$ from the
$\Phi_i$ in
order to split off a
$c=0$ theory to be left with the second potential describing the new
theory.

Since the application I have in mind is to understand string vacua the
focus in the following will not be in just any Landau--Ginzburg theory
but in special
ones, namely those defining consistent ground states. It turns out
that the
above potential $\tW$ indeed appears as part of superpotentials, but
it does so together
with other fields (in order to bring the central charge up to $c=9$).
An additional field that appears in the context of the two potentials
above is a field of charge $1/30$ to which the field of weight $11/15$
can obviously
be coupled. Thus the construction goes through. This example shows that
the framework introduced here also throws light on the concept of
the transposition of Landau--Ginzburg vacua. Since the applicability of
transpos
restricted to particular points in the moduli space it is to be
hoped that
incorporating it into the present framework allows a generalization
away from this restriction.

\vskip .3truein

\noindent
{\sc 4. Flows between Mirror Pairs of Landau--Ginzburg Theories}

\noindent
In the following I will discuss applications of the constructions of
the previous Sections to a number of examples of string vacua described
by different types of Landau--Ginzburg theories in
order to illustrate the fact that the range of applicability is
actually
quite large even though at present it is not known precisely what
class of models can be discussed in this way.

\noindent
{\it The minimal series:}

\noindent
The simplest situation is of course described by the transition from a
minimal diagonal theory to a minimal D--type polynomial. Consider
e.g. the vacuum
described by the tensor product of four minimal
N$=$2 superconformal theories $(1 6\cdot 31\cdot 86)$ endowed
with the diagonal invariant in each factor. The Landau--Ginzburg
potential
of this theory is described by the Fermat type polynomial
\beq
W= \Phi_1^3 + \Phi_2^{33} + \Phi_3^{88} +\Phi_4^8.
\eeq
The field theoretic limit of this model describes a particular point
in the
deformation class
\fnote{4}{The $*$ indicates that the GSO--projection has been
implemented.}
\beq
\IC_{(88,8,3,33,132)}^*[264]^{57}_{-48}
\eeq
the mirror of which can be shown, via the fractional transformations
introduced in [4], to be described by the potential
\beq
W= \Phi_1^3 + \Phi_2^{33} + \Phi_3^{88} + \Phi_4^4 + \Phi_2\Phi_5^2
\eeq
 This potential leads to the Calabi--Yau configuration
\beq
\IC_{(88,8,3,66,99)}^*[264]^{81}_{48}
\eeq
(here the subscripts (superscripts) denote the Euler number
(number of (1,1)--forms).The  important operation in order to produce
the mirror of the ground state
thus is to simply replace the diagonal invariant in the level 6 theory
by the D--invariant. This operation leaves the first three terms in
the potential invariant. The relevant potentials to consider are
\beq
W(\Phi_i) = \Phi_4^8+\Phi_5^2
\eeq
and
\beq
\tW(\tPhi_i) = \tPhi_4^4+\tPhi_4\tPhi_5^2
\eeq
leaving the other fields as spectators.

These two potentials, however, are just of the type
considered in eqs. (3,4) above and it is obvious that the
discussion
there immediately applies to the mirror pair of Landau--Ginzburg
theories. After including the fields
$\tPsi_i$ with negative central charge the extended theory becomes
\beq
\Phi_1^3+\Phi_2^{33}+\Phi_3^{88}+
\Phi_4^8+\Phi_5^2+\tPhi_4^4+\tPhi_4\tPhi_5^2 +
\tPhi_4\tPsi_4+\tPhi_5\tPsi_5.
\eeq
and moving to the singular configuration
\beq
\Phi_1^3+\Phi_2^{33}+\Phi_3^{88}+\tPhi_4^4+\tPhi_4\tPhi_5^2
\Phi_4^8+\Phi_5^2 + \Phi_4^2\tPsi_4 +\Phi_4^3\tPsi_5
\eeq
allows to split off the $c=0$ modification of the diagonal
configuration,
leaving the mirror theory of our starting point.
It should be noted that instead of just reinstating the monomials
defining
the original theory it is also possible to add all marginal operators
that can be constructed from the first three scaling fields, thus
allowing
us to mirror map a submanifold of the moduli space.

It is also possible to consider nondiagonal theories by adding
potentials
that do not derive from minimal N$=$2 superconformal theories. An
example
of such an application is given by the mirror pair
\beq
\IC_{(3,11,16,10,40)}^*[80]^{19}_{-16} ~\ni ~
\{\Phi_1^{23}\Phi_2 +\Phi_1\Phi_2^7 +\Phi_3^5+\Phi_4^8+\Phi_5^2=0\}
\eeq
and
\beq
\IC_{(3,11,16,20,30)}^*[80]^{27}_{16} ~\ni ~
\{\Phi_1^{23}\Phi_2 + \Phi_1\Phi_2^7 + \Phi_3^5+\Phi_4^4
+\Phi_4\Phi_5^2=0\}.
\eeq

\noindent
{\it Iterating the minimal series:}

\noindent
There exist, of course, also potentials which do not contain any
minimal factor at all and an obvious question is whether the idea I
introduced
here also applies to such theories. In fact it does, as the pair of
models
\beq
\IC_{(4,6,50,15,25}^*[100]_{-8}^{33} ~~\ni ~~
\{\Phi_1^{25} + \Phi_1\Phi_2^{16} +\Phi_3^2 + \Phi_4^5\Phi_5
+ \Phi_5^4=0\}
\eeq
and
\beq
\IC_{(4,12,44,15,25)}^*[100]_8^{37} ~~\ni ~~
\{\tPhi_1^{25}+\tPhi_1\tPhi_2^8+\tPhi_2\tPhi_3^2+
  \tPhi_4^5\tPhi_5+\tPhi_5^4=0\}
\eeq
shows which uses the potentials (\ref{minidiagit}) and
(\ref{minidit}) discussed above.

\noindent
{\it Transposed Mirror Pairs:}

An example where the transition from the original theory to the
mirror is accomplished by a simple transposition the potential is
furnished
by the following pair:
\beq
\IC_{(2,15,15,8,20)}^*[60]^{30}_{-12} ~\ni ~
\{\Phi_1^{30}+\Phi_2^4+\Phi_3^4+\Phi_4^5\Phi_5+ \Phi_5^3=0\}
\eeq
and
\beq
\IC_{(2,15,15,12,16)}^*[60]^{36}_{12} ~\ni ~
\{\Phi_1^{30}+\Phi_2^4+\Phi_3^4+\Phi_4^5 + \Phi_4\Phi_5^3=0\}
\eeq

\noindent
{\it None of the above:}

\noindent
An instructive example which involves neither the infinite series
nor the transposition procedure is provided by the pair
\beq
\IC_{(3,5,16,16,40)}^*[80]_{-8}^{33} ~~\ni ~~
\{ \Phi_1^{25}\Phi_2 +\Phi_2^{16} +\Phi_3^5+\Phi_4^5+\Phi_5^2 =0\}
\eeq
and its mirror
\beq
\IC_{(3,5,16,24,32)}^*[80]_8^{37} ~~\ni ~~
\{\tPhi_1^{25}\tPhi_2+\tPhi_2^{16}+\tPhi_3^5 + \tPhi_3\tPhi_5^2+
  \tPhi_5\tPhi_4^2=0\}.
\eeq
In this case no truncation of the potential will do in order to be
able
to decouple the $\Psi_i$'s from the original theory. Naively it
seems that
since the first part of the two potentials agrees we only need to
consider
the last three terms. Suppose we try to trivialize only the latter
part of the
second theory, i.e. we consider
\beq
\tPhi_3^5 + \tPhi_3\tPhi_5^2+
  \tPhi_5\tPhi_4^2 + \tPhi_3\Psi_3 + \tPhi_4\Psi_4 +\tPhi_5\Psi_5,
\eeq
thereby introducing fields $\Psi_i$ of weights $4/5,3/5,7/10$. If only
the fields
$\Phi_{3,4,5}$ are avaible for couplings to the $\Psi_i$ it is clear
that
the field $\Psi_3$ cannot be decoupled from the $\tPhi_i$'s.
The way out of this is by considering the remaining fields as well.

Thus adding the trivialization of the full second theory leads to
\beq
\Phi_1^{25}\Phi_2 +\Phi_2^{16} +\Phi_3^5+\Phi_4^5+\Phi_5^2 +
\tPhi_1^{25}\tPhi_2+\tPhi_2^{16} +\tPhi_3^5 +\tPhi_3\tPhi_5^2
+\tPhi_5\tPhi_4^2 +\sum_{i=1}^5 \tPhi_i \Psi_i.
\eeq
Moving to the singular configuration
\beq
\Phi_1^{25}\Phi_2+\Phi_2^{16} +\Phi_3^5+\Phi_4^5+\Phi_5^2 +
\Phi_1\Psi_1+ \Phi_2\Psi_2 +\Phi_3\Psi_3 + \Phi_3^2\Psi_4
+ \Phi_1^8 \Psi_5 +
\tPhi_1^{25}\tPhi_2+\tPhi_2^{16}+\tPhi_3^5 + \tPhi_3\tPhi_5^2+
  \tPhi_5\tPhi_4^2
\eeq
and splitting off  the trivialization of the first one leaves
the mirror.

\vskip .3truein
\noindent
{\sc 5. Constructing Mirrors via Marginal Singular Flows}

\noindent
So far I have always assumed as given a pair of theories and then
attempted to
construct a flow from one to the other. The question arises whether
it is
possible to let the construction determine to which theory the
deformed
model wants to flow. This is a somewhat involved problem which I
will discuss
only briefly in the context of the simplest theory encountered above.

Consider again the potential $\tW=\tPhi_1^4+\tPhi_1\tPhi_2^2$ with
central charge $c=9/4$ as
starting point. The fields $\Psi_i$ that appear in the trivialized
theory
have charges $3/4,5/8$ respectively. We are interested in all
potentials
$W(\Phi_1,\Phi_2)$ which satisfy the following constraints:
\begin{itemize}
\item  $W(\Phi_1,\Phi_2)$ describes a theory of central charge
$c=9/4$
\item  Marginal operators of the form $\Phi_i^{a_i}\Psi_j$
exist
\item  $W(\Phi_1,\Phi_2)$ has an isolated singularity at the
origin.
\end{itemize}
In the case at hand there are only a few possible operators that
can appear. (I) The operators can be of the form
$\Phi_1^{a_1} \Psi_1, \Phi_2^{a_2} \Psi_2$. The central charge
condition
then dictates that the only possible solution is $a_1=1=a_2$ which
are the operators used to trivialize the original theory.
(II) The operators are of the form
$\Phi_1^{a_1} \Psi_1,~\Phi_1^{a_2} \Psi_2$,
in which case $2a_2=3a_1$. Thus $a_1$ must be even $a_1=2n$ and
$q_1=1/8n$ and $q_2=\frac{1}{8}(5-\frac{1}{n})$. The requirement
that
the potential $W$ has an isolated singularity only at the origin
finally determines $n=1$ and hence we have {\it derived} with the
trivialized theory (24) the potential (23) dressed up with a
trivial factor.

\vskip .3truein

\noindent
{\sc 6. Conclusion}

\noindent
RG flows between different vacua described by conformal field
theories
at some fixed central charge (e.g. $c=9$ theories in the case of
left--right symmetric compactifications of the heterotic string)
can proceed only via marginal operators. Marginal deformations
do not, generically, change the spectrum of the theory but instead
change the Yukawa couplings and the symmetries of the theory under
consideration.

It has been shown in this article that it {\it is} possible to
flow between Landau--Ginzburg vacua with different spectra via
marginal deformations. The essential new ingredient are composite
marginal operators that lead to a singular theory, in the sense
that the chiral ring becomes ill--defined. By splitting off singular
subtheories a regular theory is obtained. The existence of such
composite marginal is nontrivial, hence the flows do not connect
 arbitrary pairs of Landau--Ginzburg theories.
The results presented here show  that the class of potentials which
is amenable to this new construction
contains polynomials that occur in mirror pairs of string vacua
described
via Landau--Ginzburg theories. An immediate consequence of these
flows therefore is a novel type of mirror map which proceeds
via singular marginal flows.

The general picture that emerges is one of a space
of trivialized Landau--Ginzburg theories of vanishing central charge
in which particular subspaces describe regular Landau--Ginzburg
at some fixed central charge c$\neq $0. These regular theories are
obtained
by splitting off large parts of the moduli space which describe
theories with vanishing central charge.

\vskip .3truein

\noindent
{\sc Acknowledgement}

\noindent
Part of this work was done while I was visiting CERN. I'm grateful
to this institution for hospitality and to the theorists there
for discussions,
in particular Per Berglund, Wolfgang Lerche and Jan Louis.

\vskip .3truein

\noindent
{\sc References}
\begin{enumerate}
\item A.B.Zamolodchikov, JETP Lett. {\bf 43}(1986)730;
      Sov.J.Nucl.Phys. {\bf 46}(1987)1090
\item P.Candelas, A.Dale, C.A.L\"utken and R.Schimmrigk,
      Nucl.Phys. {\bf B298}(1988)493
\item P.Candelas, M.Lynker and R.Schimmrigk,
Nucl.Phys. {\bf B341}(1990)383
\item M.Lynker and R.Schimmrigk, Phys.Lett. {\bf 249B}(1990)237
\item B.R.Greene and M.R.Plesser, Nucl.Phys. {\bf B338}(1990)15
\item P.Berglund and T.H\"ubsch,
        {\it A Generalized Construction of Mirror Manifolds},
        CERN preprint CERN--TH--6341/91
\item V.V.Batyrev, {\it Dual Polyhedra and Mirror Symmetry for
       Calabi--Yau Hypersurfaces in Toric Varieties}, University of
       Essen preprint
\item R.Schimmrigk, unpublished
\item E.Martinec, Phys.Lett. {\bf 217B}(1989)431
\item C.Vafa and N.Warner, Phys.Lett. {\bf 218B}(1989)51
\item Y.Kazama and H.Suzuki, Nucl.Phys. {\bf B321}(1989)232
\item W.Lerche, C.Vafa and N.Warner, Nucl.Phys. {\bf B324}(1989)427
\item D.Gepner, Nucl.Phys. {\bf B322}(1989)65
\item A.Klemm and R.Schimmrigk, {\it Landau--Ginzburg String Vacua},
CERN preprint CERN--TH--6459/92 and Heidelberg preprint
HD--THEP--92--13, to appear in Nucl.Phys. {\bf B}
\item M.Kreuzer and H.Skarke, {\it No Mirror Symmetry among
Landau--Ginzburg Vacua}, CERN preprint CERN--TH--6461/92,
to appear in Nucl.Phys. {\bf B}

\end{enumerate}

\end{document}